\documentstyle[aps]{revtex}  
\def\Paris{Par\'\i{}s}
\def\Perez{P\'erez}
\def\Astrofisica{Astrof\'\i{}sica}

\def\Fisica{F\'\i{}sica}
\def\Torrejon{Torrej\'on}
\begin{document} 
\twocolumn
\title{ 
Small-scale properties of the KPZ equation and dynamical symmetry breaking
} 
\author{ 
David Hochberg$^{+,*}$, Carmen Molina--\Paris$^{++,*}$,  
Juan \Perez--Mercader$^{+,*}$, and Matt Visser$^{+++}$ 
} 
\address{ 
$^{+}$Laboratorio de \Astrofisica\ Espacial y \Fisica\ 
Fundamental, Apartado 50727, 28080 Madrid, Spain\\ 
$^{++}$Theoretical Division, Los Alamos National Laboratory, 
Los Alamos, New Mexico 87545, USA\\ 
$^{+++}$Physics Department, Washington University, 
Saint Louis, Missouri 63130-4899, USA\\  
$^{*}$Centro de Astrobiolog\'\i a, 
INTA-CSIC, Ctra. Ajalvir, Km. 4, 28850 \Torrejon, Madrid, Spain}
\date{Revised April 2000; \LaTeX-ed \today}
\maketitle 

\begin{abstract}

A functional integral technique is used to study the ultraviolet or
short distance properties of the Kardar-Parisi-Zhang (KPZ) equation with
white Gaussian noise.  We apply this technique to calculate the one-loop
effective potential for the KPZ equation.  The effective potential is (at
least) one-loop ultraviolet renormalizable in 1, 2, and 3 space dimensions,
but non-renormalizable in 4 or higher space dimensions. This potential is
intimately related to the probability distribution function (PDF) for the
spacetime averaged field.  For the restricted class of field configurations
considered here, the KPZ equation exhibits dynamical symmetry breaking
(DSB) via an analog of the Coleman-Weinberg mechanism in 1 and 2 space
dimensions, but not in 3 space dimensions.

\bigskip


\bigskip


\bigskip

\end{abstract}

\newcommand{\Str}{\mathop{\mathrm{Str}}} 
\newcommand{\tr}{\mathop{\mathrm{tr}}} 
\newcommand{\define}{\mathop{\stackrel{\rm def}{=}}}
\newcommand{\Tr}{\mathop{\mathrm{Tr}}}
\def\d{{\mathrm d}}
\def\implies{\Rightarrow}
\def\dirac{\gamma^\mu (\partial_\mu - A_\mu)}
\def\half{ {\scriptstyle{1\over2}} }
\def\A{ {\cal A} }

In understanding the onset of spatio-temporal pattern formation in systems
out of equilibrium, it has proven extremely useful to begin the pattern
formation and selection analysis by first obtaining and then classifying
all the static and spatially homogeneous states allowed by the
time dependent partial differential equations that model the system in
question~\cite{Cross-Hohenberg}.  In this way, one can decide whether the
system will exhibit Hopf bifurcations and/or Turing instabilities and get a
handle on the qualitative nature of the patterns expected to emerge. This
can be followed up by an amplitude analysis of the {\em fluctuations} about
these static and homogeneous configurations. The unstable modes are the
ones that lead to non-trivial patterns.  For out-of-equilibrium systems
coupled to noisy environments (or with inherent internal noise) it is
important to know how the stochastic sources can alter and shift these
static and homogeneous configurations, since these affect the onset of the
pattern-forming instabilities.  Here we show how the effects of noise on
these configurations can be computed using functional integral methods, and
reveal an intrinsically ultraviolet one-loop phenomenon not captured by the
more common methods of analysis described above: namely the occurrence of
dimension dependent dynamical symmetry breaking. In this paper we focus
attention on the Kardar-Parisi-Zhang (KPZ) equation~\cite{KPZ,MHKZ}
\begin{equation}\label{KPZ}
\left({\partial\over\partial t} - \nu \vec\nabla^2\right) \phi = 
F_0+{\lambda\over2} (\vec\nabla\phi)^2 + \eta.
\end{equation}
The KPZ equation arises in a host of seemingly distinct physical contexts
and problems ranging from models of turbulence~\cite{FNS,Frisch}, interface
growth, driven diffusion and flame fronts~\cite{MHKZ}, directed polymers
in a random medium~\cite{Lassig,Bundschuh}, certain lattice gases with
hard-core exclusion~\cite{Derrida}, and even structure development in the
early universe~\cite{Structure}.  In the fluid dynamics interpretation
(when the KPZ equation is used as a model of Burgers turbulence), the fluid
velocity is $\vec v = - \vec\nabla \phi$ and the KPZ field plays the r\^ole
of a velocity potential.  In the surface growth interpretation $\phi(\vec
x,t)$ is the local height of the surface, defined over a two-dimensional
plane~\cite{KPZ}. The constant ``tadpole'' term $F_0$ in (\ref{KPZ}) is
necessary for the ultraviolet regularization~\cite{HMPV-spde} of the KPZ
equation\footnote{In this regard it is interesting to point out that small
scale properties of a randomly stirred fluid were studied some years ago by
Yakhot, who found that the ultraviolet renormalizability of the forced
Navier-Stokes (NS) equation required the addition of a term not originally
present in the bare NS equation~\cite{Yakhot}.}.  After renormalization, we
will argue that the tadpole can and should be set to zero.

If the noise $\eta$ is Gaussian, all stochastic averages are encoded in the
generating functional\footnote{This ``direct'' functional integral
formalism is different (yet equivalent) from the more extended MSR
approach~\cite{MSR}.}~\cite{HMPV-spde}
\begin{eqnarray}\label{functional}
Z[J] &=& 
\int ({\cal D} \phi)\;
\exp\left( \int J \phi \right)
\nonumber\\
&&
\exp\Bigg( 
-{1\over2} \int \int
\left[
\partial_t \phi - \nu\vec\nabla^2 \phi 
-F_0 - {\lambda\over2} (\vec\nabla\phi)^2
\right]
\nonumber\\
&& 
\qquad
G_\eta^{-1}
\left[
\partial_t \phi - \nu\vec\nabla^2 \phi 
-F_0 - {\lambda\over2} (\vec\nabla\phi)^2
\right]
\Bigg).
\end{eqnarray}
In general, a functional Jacobian determinant must be included. For the KPZ
equation this determinant is a field-independent constant~\cite{HMPV-spde}.
For translation-invariant Gaussian noise, we split its two-point function
into an {\em amplitude}, $\A$, and a {\em shape},
$g_2(x,y)$~\cite{HMPV-spde}
\begin{equation}
G_\eta(x,y) \define \A \; g_2(x-y),
\end{equation}
with the {\em convention} that
\begin{equation}
\int g_2^{-1}(\vec x,t)\; \d^d \vec x\; \d t 
\; = \; 1
\; = \; \tilde g_2^{-1}(\vec k=\vec 0,\omega=0).
\end{equation}
The amplitude $\A$ is the loop-counting parameter for this
theory~\cite{HMPV-spde}. 

There are two important symmetries of the KPZ equation that are
relevant for our analysis. First, we have
\begin{eqnarray}
\phi &\to& \phi + c(t), \\
F_0  &\to& F_0 + {d c(t)\over dt}. 
\end{eqnarray}
In the fluid dynamics interpretation this symmetry is a ``gauge
transformation'' of the scalar field $\phi$ that does not change the fluid
velocity $\vec v$. In the surface growth interpretation this symmetry is a
(Type I) Galilean transformation. It can be used to eliminate any spurious
motion of the background field. The second symmetry is ($\vert \epsilon \ll
1 \vert$)
\begin{eqnarray}
\vec x &\to& \vec x' = \vec x - \lambda \; \vec \epsilon \; t, \\
t &\to& t' = t, \\
\phi(\vec x,t) &\to& \phi'(\vec x',t') 
= \phi(\vec x,t) - \vec \epsilon \cdot \vec x.
\end{eqnarray}
In the fluid dynamics interpretation this symmetry is equivalent to a
(Type II) Galilean transformation of the fluid velocities
\begin{equation}
\vec v \to \vec v' = \vec v + \vec \epsilon
\; .
\end{equation}
In the surface growth interpretation this symmetry amounts to {\em tilting}
the coordinate system at an angle to the vertical, with
\begin{equation}
\tan(\theta) = \vert \vert \,\vec \epsilon\, \vert \vert
\; .
\end{equation}
This Type II Galilean transformation is an exact invariance of the
zero-noise KPZ equation, but once noise is added to the system, it will
remain a symmetry only if the noise is translation-invariant and {\em
temporally white}.

A few words are in order regarding the choice of field configurations used
for calculating the effective potential. In quantum field theory (QFT)
patterns of spontaneous and dynamical symmetry breaking are revealed using
static and homogeneous fields, since these configurations minimize the
positive kinetic energy contribution to the total energy. The QFT
ground states (basic field configurations) are therefore determined by
searching for the minima of the effective potential.  It turns out that an
effective potential for the KPZ equation can be defined for a static, but
spatially inhomogeneous field configuration, if it has the form
\begin{equation}\label{ramp}
\phi = - \vec v \cdot \vec x.
\end{equation}
In the hydrodynamic interpretation this corresponds to a stationary and
homogeneous fluid flow~\cite{Frisch} with velocity $\vec v$.  In the
surface growth interpretation $\vert \vert \vec v \vert \vert$ corresponds
to a constant slope of the surface~\cite{KPZ} (linear ramp). Due to the
structure of the stochastic field theory (\ref{functional}) leading to the
KPZ equation, this is the most general {\em static} configuration allowing
one to define an effective potential.  This is because the effective
potential must be a function of a static field that satisfies the heat
equation: $\partial_t \phi - \nu \nabla^2 \phi = 0$.  Any other field
configuration more general than (\ref{ramp}) (and/or time dependent) would
force us to deal with the complexities of the effective action.

Following the general construction in~\cite{HMPV-spde}, the KPZ
zero-loop effective potential is given by

\begin{equation}\label{barepotential}
{\cal V}_{\mathrm zero-loop}[v] = 
\half \left[\left( F_0 + \half{\lambda} v^2 \right)^2 - F_0^2\right].
\end{equation}
The effective potential is generally calculated in terms of a reference
background field $\phi_0 = -{\vec v}_0 \cdot {\vec x}$. We have used the
Type II symmetry to set $\vec v_0 = \vec 0$. This potential is formally
equivalent to that of $\lambda\phi^4$ QFT -- with the velocity $\vec v$
playing the role of the quantum field $\phi_{\mathrm QFT}$. Even at
zero-loops we see that for $F_0<0$, the effective potential takes on the
``Mexican hat'' form, so that the onset of spontaneous symmetry breaking
(SSB) would not be at all unexpected~\cite{Weinberg,Zinn-Justin}. However,
because of the Type I symmetry, the renormalized value of $F_0$ must be set
to zero, and SSB is not encountered in the KPZ equation.  (There is an
instructive analogy with the renormalization program for massless $\lambda
\phi^4$ QFT.  It is well known that for this QFT one still needs to
introduce a bare mass parameter that must be kept for all {\em intermediate
stages} of the regularization. It is only {\em after} renormalization is
complete that it makes sense to set the renormalized value of the mass to
zero.)  Considerably more subtle is the onset of DSB which can be detected
only via a one-loop computation.

Evaluating the general expression for the one-loop effective potential
computed in~\cite{HMPV-spde} for the case of the KPZ equation, we obtain
\begin{eqnarray}
{\cal V}[v] &=& 
\half(F_0+\half{\lambda} v^2)^2
\nonumber\\
&+&
\half \A \int {\d^d \vec k \; \d \omega\over (2\pi)^{d+1}}
\log  
\left[ 1 + {\tilde g_2(\vec k,\omega) 
\lambda (F_0+\half{\lambda} v^2) \vec k^2
\over 
(\omega - \lambda \vec v \cdot \vec k)^2 + \nu^2 (\vec k^2)^2}
\right]
\nonumber\\
&-&
 \left( \vec v \to \vec 0 \right)
+ O(\A^2).
\end{eqnarray}
Because we have chosen noise which is {\em white} in time, so that $\tilde
g_2(\vec k,\omega)\to \tilde g_2(\vec k)$, we can shift the integration
variable from $\omega$ to $\omega - \lambda \vec v \cdot \vec k$. The
resulting integral becomes
\begin{eqnarray}
{\cal V}[v] &=& 
\half\left[ (F_0+\half{\lambda} v^2)^2 - F_0^2 \right]
+ \half \A \int {\d^d \vec k \; \d \omega\over (2\pi)^{d+1}}
\nonumber\\
&\times&
\log  
\left[ {\omega^2 + \nu^2 (\vec k^2)^2 
+\tilde g_2(\vec k) 
\lambda (F_0+\half{\lambda} v^2) \vec k^2
\over 
\omega^2 + \nu^2 (\vec k^2)^2 
+\tilde g_2(\vec k) 
\lambda F_0 \vec k^2}
\right]
\nonumber\\
&+&
 O(\A^2).
\nonumber
\end{eqnarray}
We point out that the denominator of the logarithm in the integrand is
independent of the velocity field $\vec v$. This important result is a
consequence of the fact that the Jacobian functional determinant
encountered in~\cite{HMPV-spde} is constant for the KPZ equation. One can
perform the frequency integral exactly to obtain
\begin{eqnarray}
\label{one-loop}
{\cal V}[v] &=& 
\half \left[ (F_0 + \half{\lambda} v^2)^2 - F_0^2 \right]
\nonumber\\
&+&
\half \A \int {\d^d \vec k  \over (2\pi)^{d}} 
\Bigg\{
\sqrt{\left[   \nu^2 (\vec k^2)^2 
+ \tilde g_2(\vec k) 
\lambda (F_0+\half{\lambda} v^2) \vec k^2
\right]} 
\nonumber\\
&-& 
 \sqrt{\left[ \nu^2 (\vec k^2)^2 
+ \tilde g_2(\vec k)
\lambda F_0 \vec k^2
\right]}
\Bigg\}
+ O(\A^2).
\end{eqnarray}
The bare KPZ potential (\ref{barepotential}) contains terms proportional to
$v^0$, $v^2$, and $v^4$. (The one-loop contribution (\ref{one-loop}),
expanded in powers of $v^2$, has terms proportional to $v^{2n}$.) Let us
now take the spatial noise spectrum to be cutoff white noise, {\em i.e.},
\begin{equation}
\tilde g_2(\vec k) 
= \tilde g_2\left(\,\vert\vert \vec k\vert\vert\,\right) 
= \Theta(\Lambda - k).
\end{equation}
With this choice of noise, the $v^2$ term is proportional to $\Lambda^d$,
the $v^4$ term to $\Lambda^{d-2}$, and the $v^6$ term to
$\Lambda^{d-4}$. In order to be able to absorb the infinities into the bare
action we must have $d<4$. Note that $d=4$ is a marginal case, and one
might suspect that $d=4$ is a critical dimension for the KPZ
equation~\cite{Lassig,Bundschuh}.  That is: the KPZ equation (subject to
white noise) is one-loop ultraviolet renormalizable {\em only} in 1, 2, and
3 spatial dimensions. Even so, one-loop renormalizability requires an
explicit tadpole. (Without a tadpole there is no term proportional to $v^2$
in the zero-loop potential, and hence no possibility of regularizing the
leading divergence.)  Strictly speaking, the claim of one-loop
renormalizability requires investigation of the effective action
(wave-function renormalization). A one-loop ultraviolet renormalization of
the effective {\em action}, valid for time dependent and inhomogeneous
fields, has been carried out which fully supports the effective potential
calculations presented here~\cite{sdw-kpz}.

The ultraviolet renormalizability of the KPZ equation depends critically on
the ultraviolet behavior of the noise. Let us suppose that the noise is
power-law distributed in the ultraviolet region with $g_2(k) \approx
(k_0/k)^{\theta} \; \Theta(\Lambda-k)$. In this case the $n$th term in the
expansion has ultraviolet behavior proportional to $v^{2n}
\Lambda^{d+2-2n-n \theta}$. The KPZ equation is then one-loop ultraviolet
renormalizable for $d<4+3\theta$.  Thus, the one-loop ultraviolet
renormalizability can be extended to dimensions four and greater depending
on how the noise scales in the ultraviolet region.  We will not pursue
these issues further in this paper and henceforth restrict our attention to
white noise.

There is a formal connection with $\lambda (\phi^4)_{[d+1]+1}$ QFT that is
made apparent by extracting a factor of $\nu^2 \vec k^2$ from equation
(\ref{one-loop}). If we do so, we obtain
\begin{eqnarray}
{\cal V}[v] &=& 
\half \left[ (F_0+\half{\lambda} v^2)^2 - F_0^2 \right]
\nonumber\\
&&
+ \half \A \nu \int_0^\Lambda {\d^d \vec k |k|  \over (2\pi)^{d}} 
\Bigg\{
\sqrt{  \vec k^2 
+  {\lambda\over\nu^2} (F_0+\half{\lambda} v^2)
}
\nonumber\\
&& \qquad 
- \sqrt{ \vec k^2 
+  {\lambda\over\nu^2} F_0 
}
\Bigg\}
+ O(\A^2).
\end{eqnarray}
This is recognizable as the effective potential for $\lambda\phi^4$ QFT in
$d+1$ spatial dimensions, that is, Euclidean $[d+1]+1$ spacetime
dimensions~\cite{Weinberg,Zinn-Justin}.  To make the connection, interpret
$\lambda F_0/\nu^2$ as the mass term $m^2$ of the QFT, $\lambda^2/\nu^2$ as
$\lambda_{\mathrm QFT}$ (the coupling constant of the QFT), and $v$ as the
field $\phi_{\mathrm QFT}$.

After renormalization the tadpole is to be set to zero, but we still need a
bare tadpole to act as a counterterm during the intermediate stages of the
regularization. This is completely analogous to the situation in massless
$\phi^4$ QFT, where the renormalized mass of the quantum field is
fine tuned to zero by hand~\cite{Weinberg,Zinn-Justin}.

We now evaluate the effective potential in different spatial dimensions.


\underline{Case $d=1$:} We are interested in the integral
\begin{equation}
{\cal I}(a) \define 
\int_0^{+\infty} \d k^2 \left[ \sqrt{k^2+ a}- \sqrt{k^2} \right].
\end{equation}
This integral is divergent and needs to be regularized. The most direct way
to do so is by the ``differentiate and integrate'' trick which leads
to~\cite{Weinberg}
\begin{equation}
{\cal I}(a) = \kappa a - {2\over3} a^{3/2}.
\end{equation}
Here $\kappa$ is an infinite constant of integration. We absorb $\kappa$
into the bare action, where it renormalizes
$F_0$~\cite{Weinberg,Zinn-Justin}
\begin{eqnarray}
\label{E:KPZ-1}
{\cal V}[v;d=1] &=& 
\half\left[ (F_0+\half{\lambda} v^2)^2 - F_0^2 \right]
\nonumber\\
&&
- {1\over6\pi} \A {\lambda^{3/2} \over \nu^2} 
\left[
\left( F_0+\half{\lambda} v^2\right)^{3/2} - F_0^{3/2}
\right]
\nonumber\\
&&
+ O(\A^2).
\end{eqnarray}
These are all renormalized parameters at $O(\A)$.  The zero-loop
contribution will always dominate at large fields (rapid flow).  Near $v =
0$, it is the one-loop contribution that is dominant.  We take the physical
value of the {\em renormalized} tadpole, $F_0=0$, and encounter something
very interesting -- the system undergoes dynamical symmetry breaking (DSB)
in a manner qualitatively similar to the Coleman-Weinberg mechanism of
QFT.  The effective potential then simplifies to
\begin{eqnarray}
{\cal V}[v;d=1] &=& 
{\lambda^2\over8}v^4
- {1\over6\pi} \A {\lambda^3 \over {2^{3/2}\nu^2}} |v|^3
+ O(\A^2).
\end{eqnarray}
The potential is not analytic at zero field (a phenomenon well known from
massless QFTs)~\cite{Weinberg,Zinn-Justin}.  For large $v$ the classical
potential dominates, whereas for small $v$ one-loop effects dominate. Thus
the symmetric vacuum ($v=0$) is unstable.  While it is easy to see that the
symmetry is dynamically broken, the presence of the unknown $O(\A^2)$ terms
make it difficult to make a quantitative estimate for $v_{\mathrm min}$.
We can nevertheless {\em qualitatively estimate} the shift in the
expectation value of the velocity field
\begin{equation}\label{vmin1}
v_{\mathrm min} = 
\pm \A \; {\lambda\over2\pi \; 2^{1/2} \; \nu^2} + O(\A^2).
\end{equation}
This DSB is particularly intriguing in that it suggests the possibility of
a noise driven pump. For example, in thin pipes (where the flow is
essentially one-dimensional) and provided the physical situation justifies
the use of Burgers equation, this result indicates the presence of a
bimodal instability leading to the onset of a fluid flow with velocity
depending on noise amplitude.


\underline{Case $d=2$:} Because of the analogy between the one-loop
effective potential for the KPZ equation (for white noise) and that for the
$\lambda\phi^4$ QFT, we can write down the renormalized one-loop effective
potential by inspection~\cite{Weinberg,Zinn-Justin}
\begin{eqnarray}
&&{\cal V}[v;d=2] = 
\half\left\{ [F_0(\mu)+\half{\lambda(\mu)} v^2]^2 - [F_0(\mu)]^2 \right\} 
\nonumber\\
&&
\qquad
+\half \A  {1  \over (2\pi)^{2}} 
{\lambda^2\over\nu^3}
\Bigg\{
[F_0(\mu)+\half{\lambda(\mu)} v^2]^2 
\nonumber\\
&&
\qquad
\times
\log \left[{F_0(\mu)+\half{\lambda(\mu)} v^2\over \mu^2 }\right]
-[F_0(\mu)]^2 
\log \left[{F_0(\mu)\over \mu^2 }\right]
\Bigg\}
\nonumber\\
&&
\qquad
+ O(\A^2).
\end{eqnarray}
Here $\mu$ is the renormalization scale in the sense it is used in
QFT~\cite{Weinberg,Zinn-Justin}.

We now tune the renormalized tadpole to its physical value of zero.  After
performing a finite renormalization to simplify the expression, we obtain
\begin{eqnarray}
{\cal V}[v;d=2] &=& 
{\lambda^2\over 8} v^4 
+\A  {1  \over (2\pi)^{2}} 
{\lambda^4\over8\nu^3}
v^4 
\log \left({v^2\over \mu^2 }\right)
+ O(\A^2).
\nonumber\\
&&
\end{eqnarray}
The one-loop contribution will always dominate at large fields (rapid
flow). Near $v = 0$ the logarithmic singularity is rendered finite by the
polynomial prefactor.  That the KPZ potential has a non-trivial minimum
exhibiting DSB is exactly the analog of the Coleman-Weinberg
mechanism~\cite{Weinberg,Zinn-Justin}.  We can estimate the location of the
minimum by differentiating the effective potential. We obtain
\begin{equation}\label{vmin2}
v_{\mathrm min} = \pm \mu \;
\exp\left[
-{(2\pi)^2\;\nu^3\over2\;\lambda^2 \;\A} - {1\over4}  +O(\A) 
\right].
\end{equation}
Note that $v_{\rm min} \to 0$ as $\A \to 0$, to recover the tree level
minimum $v_{\rm min}=0$.  The present discussion is relevant to either (1)
surface evolution on a two dimensional substrate, or (2) thin superfluid
films, since superfluids are automatically vorticity free, justifying the
application of the KPZ equation.

We can also calculate the one-loop beta function, for as the bare effective
potential does not depend on the renormalization scale, one has that $ {\mu
\; \d {\cal V}/ \d\mu} =0$. We get
\begin{equation}
\beta_\lambda \define \mu \; {\d \over \d\mu} \; \lambda = 
{\A \over4\pi^2} {\lambda^2\over\nu^3} \lambda+ O(\A)^2
\; .
\end{equation}
%


\underline{Case $d=3$:} We are interested in the integral
\begin{equation}
{\cal I}(a^2) \define 
\int_0^{+\infty} \d k^2 k^2 \left[ \sqrt{k^2+ a^2}- \sqrt{k^2} \right].
\end{equation}
The ``differentiate and integrate'' trick yields~\cite{Weinberg}
\begin{equation}
{\cal I}(a^2) = \kappa_1 a^2 +\kappa_2 a^4 + {4\over15} a^{5}.
\end{equation}
Here $\kappa_1$ and $\kappa_2$ are two infinite constants of integration.
Since the sign in front of the $a^5$ term is positive there is no
possibility of DSB in $d=3$.  We absorb $\kappa_1$ and $\kappa_2$ into the
bare action, where they renormalize both $F_0$ and $\lambda$
\begin{eqnarray}
{\cal V}[v;d=3] &=& 
\half\left[ (F_0+\half{\lambda} v^2)^2 - F_0^2 \right] 
\nonumber\\
&&
+ {1\over (30\pi^2)}\A {\lambda^{5/2} \over \nu^4}
\left[
\left( F_0+\half{\lambda} v^2\right)^{5/2} - F_0^{5/2}
\right]
\nonumber\\
&&
+ O(\A^2).
\end{eqnarray}
These are all renormalized parameters at order $O(\A)$.  For large fields
the one-loop correction is dominant, whereas for small fields the tree
level result dominates. The effective potential is positive and
monotonically increasing.

If we set the renormalized tadpole to its physical value of zero, we have
\begin{eqnarray}
{\cal V}[v;d=3] &=& 
{\lambda^2\over8} v^4
+ {1\over (30\pi^2)}\A {\lambda^{5} \over 2^{5/2} \nu^4} |v|^5
+ O(\A^2).
\end{eqnarray}
At zero-loops the vacuum is symmetric at $v_{\rm min}=0$.  One-loop physics
does not change this. There is an important sign change in the one-loop
contribution when comparing $d=3$ with $d=1$. The DSB that is so
interesting in $d=1$ and $d=2$, is now absent in $d=3$. Notice that the
effective potential is non-analytic at $v_{\rm min}=0$. This case is of
most interest in hydrodynamics and we have shown that (at one-loop) the
ground state ($v_{\rm min}=0$) is stable under Gaussian noise.


\underline{Discussion:} When taken as a model for turbulence, it is natural
to study the ultraviolet properties of the KPZ equation because in a
turbulent fluid, energy cascades down from larger to smaller length scales
and one might expect scale-invariant solutions in the limit as $k
\rightarrow +\infty$. Indeed, Forster, Nelson, and Stephen pointed out long
ago the importance of studying the short distance and short time (i.e.,
ultraviolet) correlations in the fluid velocity in fully developed
turbulence~\cite{FNS}. More recently, Yakhot~\cite{Yakhot} undertook an
analysis of the ultraviolet properties of the randomly forced Navier-Stokes
(NS) equation and found that ultraviolet renormalizability requires
modifying the fluid dynamical equation by a term {\em not originally
present} in the NS equation, though this new term is much more complicated
than the tadpole needed to successfully renormalize the KPZ equation.

The ultraviolet properties of the KPZ equation exposed in this paper reveal
a rich structure and complements studies concerned with the long distance
(infrared) features of this equation.  This distinction shows up in a
striking way in both, the need of a bare tadpole for ultraviolet
renormalizability and the presence of dimension dependent dynamical
symmetry breaking for the configurations studied here.  Our focus in this
paper has been on the concept of the effective potential and we have shown
how this concept, so useful in QFT, can be extended to stochastic equations
containing gradient forcing terms, such as the KPZ equation, by employing
inhomogeneous and static fields.

The evidence for the dynamical symmetry breaking found here must be
interpreted with due caution. We must emphasize, for the reasons explained
earlier, that the effective potential is calculated for a restricted class
of field configurations and the symmetry breaking is with respect to this
class of constant fluid velocity configurations. Thus, while for $d=3$ the
one-loop effective potential is minimized for a zero-velocity background
flow $v_{\rm min} = 0$, the potential in both $d=1$ and $d=2$ is minimized
at a non-zero value of the fluid flow, namely (\ref{vmin1}) and
(\ref{vmin2}), respectively.  The symmetry breaking is thus manifested as a
preference for the system to ``jump'' from zero to a non-zero but constant
flow. The jump is triggered, of course, by the noise source.  In the
surface growth interpretation, the minimum of the potential for $d=3$
corresponds to a surface with zero slope, while the dynamical symmetry
breaking that occurs in $d=2$ and $d=1$ corresponds to a jump in the
surface slope to a constant non-zero value.  This should be contrasted to
recent exact and numerical calculations of the PDF for the KPZ equation in
$d=1$ dimension. By making use of a lattice version of the KPZ equation in
one dimension, Derrida and Lebowitz have performed an exact calculation of
the probability distribution of the deviation of the average
current~\cite{Derrida}.  The results of these authors agree with a
numerical calculation of the KPZ PDF in $d=1$ by Yakhot and
Chekhlov~\cite{Yakhot-Chekhlov}.  In these papers, the PDF, which is
essentially the exponential of the effective potential, gives information
about the probability to find a velocity difference $\Delta u_r = v(x + r)
- v(x)$ at two widely separated points in the fluid (i.e., as $r
\rightarrow +\infty$) and for asymptotically large times. Their PDFs have a
global maximum at $\Delta u_r = 0$, corresponding to no velocity
difference, i.e., a spatially constant velocity (or a constant slope
surface). Thus, the most probable configuration corresponds to a spatially
constant velocity.  Of course, these PDFs take into account {\em all}
dynamically allowed configurations that evolve via the KPZ equation; our
PDF, by contrast, samples {\em only the class} of constant velocity
configurations. Nevertheless, and within these limitations, we have been
able to show that for the linear theory ($\lambda = 0$) the ground state
has zero velocity (or zero slope) and that once we turn on the nonlinear
interaction, and restrict the fluid to have a constant velocity, there is a
dynamically preferred velocity as determined by one-loop physics in $d=1$
and $d=2$ dimensions.
   
In this paper we have also demonstrated that the KPZ equation with white
noise is at least one-loop renormalizable for $d=1,2,3$ dimensions but is
non-renormalizable (at one-loop) for $d > 4$. In making statements about
renormalizability, care must be taken in distinguishing ultraviolet and
infrared properties. Thus, from the work of Sun and
Plischke~\cite{Sun-Plischke} and Frey and T\"auber~\cite{Frey-Tauber}, it
is known that the KPZ equation is, at least, two-loop renormalizable for
$d=1,2,3$. Although these authors were interested in the infrared
properties of the KPZ equation, they had to deal with both ultraviolet and
infrared divergences in their work. By making use of the mapping of the KPZ
equation to directed polymers with quenched noise, the one-loop
renormalizability has been demonstrated in $2 < d < 4$~\cite{Bundschuh}.

Though the calculations presented here are limited to one-loop order in the
noise amplitude, it must be emphasized that in many cases one-loop physics
is enough to capture essential features of stochastic partial differential
equations (SPDEs)~\cite{HMPV-spde} and QFTs. Furthermore, while
calculations of the one-loop effective potential are straightforward, one
should not forget that it is necessary to develop a physical interpretation
of this effective potential to ensure that it is as relevant for SPDEs, as
it is for QFTs~\cite{HMPV-spde}.  In this paper we have presented an
analogy between the statics of the KPZ equation and the static behavior of
$\lambda\phi^4$ QFT --the vacuum structure of the $\phi^4$ QFT carries over
into the basic field configurations of the KPZ equation. In 1 and 2 space
dimensions we have exhibited the occurrence of DSB, which does not take
place in 3 space dimensions.

\end{document}